# Derivation of the Gutenberg-Richter Empirical Formula from the Solution of the Generalized Logistic Equation

Lev A. Maslov[1], Vladimir M. Anokhin[2]

[1] Aims College, Greeley, CO, USA lev.maslov@aims.edu, and Computer Center FEBRAS, Russia ms_leo@hotmail.com

[2] All-Russia Gramberg Research Institute for Geology and Mineral Resources of the Ocean (VNIIOKEANGEOLOGIA), St Petersburg, Russia vladanokhin@yandex.ru

**Abstract**

We have written a new equation to study the statistics of earthquake distributions. We call this equation "the generalized logistic equation".

The Gutenberg-Richter frequency-magnitude formula was derived from the solution of the generalized logistic equation as an asymptotic case for the approximation of large magnitudes. To illustrate how the solution of the generalized logistic equation works, it was used to approximate the observed cumulative distribution of earthquakes in four different geological provinces: the Central Atlantic (40N-25N, 5W-35W), Canary Islands, Magellan Mountains (20N-9S, 148E-170E), and the Sea of Japan. This approximation showed an excellent correlation between the theoretical curves and observed data for earthquakes of magnitudes *1<m<8*.

**Introduction**

Earthquakes and volcanic eruptions are very complex phenomena and are caused by many different processes taking place both inside the Earth and very far outside its outer boundaries. Many of these processes are very poorly understood. In this case, a phenomenological study of regularities in the distribution of earthquakes and faults, based on well formulated and mathematically expressed physical models, can provide insight into the physics and kinetics of earthquake processes. An example of this approach is the M. Ishimoto and K. Iida (1939) *empirical* relation between the magnitude and frequency of earthquakes, which initiated the study of earthquakes as self-organized critical phenomena. The empirical expression, similar to that obtained by M. Ishimoto and K. Iida, was published by R. Gutenberg and C.F. Richter three years later, (1942). In the current research we traditionally refer to this relation as the "Gutenberg-Richter magnitude-frequency formula": $log_{10}(F) = a - b \cdot m$. Here *F* represents frequency, *m* represents magnitude, and *a* and *b* are constants. The discovery of this formula sparked the study of earthquakes in various fields including statistical physics, fractal mathematics, and mathematics of self-organized criticality (Main, 1996), which continues to this day (Bundle, et al., 2003). The other, more technical problem, attracting the attention of scientists, is the study of the spacial and temporal variation of the coefficient *b*. It was found in numerous studies (Ayele, Kulhanek, 1997; Gerstenberger, Wiemer, Gardini, 2001; Gibowicz, Lasocki, 2001; Smyth, Mori, 2009) that this coefficient, supposedly a fundamental seismologics constant, varies significantly in space and time. It was shown also that the linear relation holds only for magnitudes *m* in a certain range, approximately $4 \leq m$. This implies that the Gutenberg-Richter magnitude-frequency empirical formula is not a "fundamental law", but part of a more general relation between the frequency of earthquakes and their magnitudes (or seismic moment).

The main objective of the present work is to derive the Gutenberg-Richter magnitude-frequency empirical formula from the solution of the generalized logistic equation and to illustrate its usage by modeling cumulative earthquake distributions in four different geological provinces.

**The generalized logistic equation**

Let us consider the equation:

$$\frac{dG(x)}{dx} = r \cdot [M - G(x)] \cdot G(x) \cdot x^{\alpha} \qquad (1)$$

with respect to the function $G(x), x \geq 0$, where $r > 0, M > 0, \alpha \in R$, are some parameters. This equation differs from the classical logistic equation only by the multiple $x^{\alpha}$.

Let

$$F = G(x)/M, s = r \cdot M,$$

then, equation (1) becomes

$$\frac{dF}{dx} = s \cdot (1-F) \cdot F \cdot x^{\alpha}. \qquad (1a)$$

The boundary condition is

$$F(x_0) = F_0. \qquad (1b)$$

The solution to the problem (1a), (1b) for $\alpha \neq -1$ is:

$$F(x) = \begin{cases} \left(1 + A\exp\left(-\frac{s}{\alpha+1} x^{\alpha+1}\right)\right)^{-1}, x > 0, \\ 0, x \leq 0. \end{cases} \qquad (2)$$

For $\alpha = -1$, the solution of (1a) and (1b) is

$$F(x) = \begin{cases} (1 + Ax^{-s})^{-1}, x > 0, \\ 0, x \leq 0. \end{cases} \qquad (3)$$

and for $\alpha = 0$ in (2) we have

$$F(x) = \begin{cases} (1 + Ae^{-sx})^{-1}, x > 0, \\ 0, x \leq 0. \end{cases} \qquad (4)$$

The constant $A$ in (2), (3) and (4) can be found using the boundary condition (1b).
The derivative of (2) for $x > 0$ is

$$\frac{dF}{dx} \equiv f(x) = \frac{s \cdot x^{\alpha} \cdot A\, exp\left(-\frac{s}{\alpha+1} x^{\alpha+1}\right)}{\left(1 + A\, exp\left(-\frac{s}{\alpha+1} x^{\alpha+1}\right)\right)^2} \quad (5)$$

Thus, the solutions of equation (1a) for $-1 \leq \alpha \leq 0$ are a set of functions distributed from log-logistic to logistic.

**The Gutenberg-Richter formula**

The Gutenberg-Richter relation between the number of earthquakes with magnitudes $\geq m$ written for the natural logarithm is:

$$ln(F^*) = a_e - b_e \cdot m \quad (6)$$

Where $F$ is the frequency, and $m$ is the magnitude of earthquakes. This formula can be rewritten as

$$F^*(m) = exp\left(-b_e\left(m - \frac{a_e}{b_e}\right)\right) . \quad (7)$$

which is the well-known exponential distribution shifted by the value of $\frac{a_e}{b_e}$.

The same expression as (7), but for magnitudes $\leq m$ has the form

$$F(m) = 1 - F^*(m). \quad (8)$$

Thus, (6, 7, 8) are equivalent forms of the Gutenberg-Richter formula. An example of this approximation of the earthquake data by an exponential distribution is shown in Figure 1.

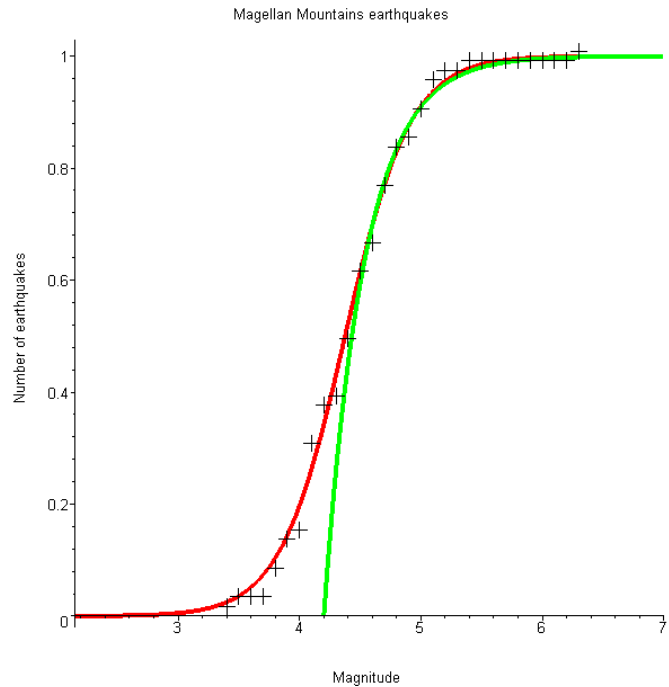

**Figure 1.**

The exponential distribution function calculated by formula (8), green; solution (2) of equation (1a), red, and observed Magellan Mountains cumulative earthquake distributions, crosses. The Gutenberg-Richter *b-value*, calculated from approximating exponential curve is *1.3*.

This figure clearly shows that the Gutenberg-Richter formula is the part of the more general law relating earthquake frequency and magnitude. This graph works in favor of the conclusion made by many authors, (Ayele, Kulhanek, 1997; Gerstenberger, Wiemer, Gardini, 2001; Gibowicz, Lasocki, 2001), that the best correlation between the Gutenberg-Richter relation and earthquake distributions is observed for large magnitudes, approximately greater than *4.0*.

**Derivation of the Gutenberg-Richter formula from the solution of the generalized logistic equation**

To compare (2) and (7) let us rewrite (2) in a form (8) for $x > 0$

$$F(m) = \frac{A \exp\left(-\frac{s}{\alpha+1} m^{\alpha+1}\right)}{1 + A \exp\left(-\frac{s}{\alpha+1} m^{\alpha+1}\right)}. \quad (9)$$

Using an expansion of (9) in powers of $A e^{-\frac{s}{\alpha+1} m^{\alpha+1}}$ we obtain

$$F(m) \approx A e^{-\frac{s}{\alpha+1} m^{\alpha+1}}, \quad (10)$$

and

$$\ln F(m) \approx \ln A - \frac{s}{\alpha+1} m^{\alpha+1} \quad (11)$$

Putting in (11) $\alpha = 0, a_e = \ln A, b_e = s,$ we arrive at the Gutenberg-Richter formula (6a). Expression (11) can be considered as a generalized Gutenberg-Richter formula.

**Application of the solution of the generalized logistic equation for modeling cumulative distribution of earthquakes**

To illustrate the application of solution (2) for modeling the cumulative distribution of earthquakes, we picked four different geological provinces: Central Atlantic, Canary Islands, Magellan Mountains, and the Sea of Japan. In addition to solution (2), the following statistical functions were used to approximate the observed data: the Gamma-distribution, the Weibull distribution, the log-normal, logistic, and log-logistic distributions. Figure 2 shows the observed cumulative normalized distribution of earthquakes with magnitudes $\leq m$ (crosses), and graphs of functions calculated from formulas (2) and (5): cumulative distribution (red), and the probability density function (blue).

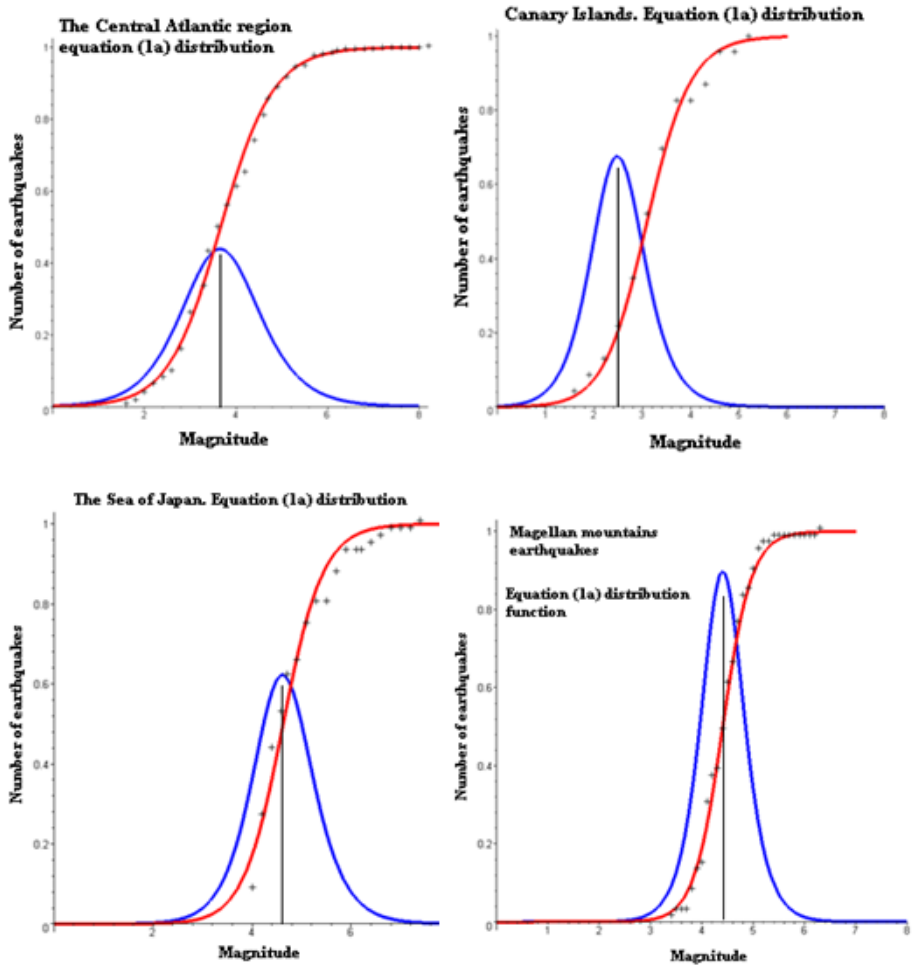

**Figure 2.**

The observed cumulative normalized distribution of earthquakes (crosses), theoretical cumulative distribution (red), and probability density function (blue).

The accuracy of our approximation was estimated by using an $\chi^2$ test for accuracy of fit with confidence level 0.05. It was found that the best approximation of the observed data for magnitudes $1 < m < 8$ was provided with distribution (2), and as in case of the Central Atlantic for $1 < m < 9$. Future research can specify the limits of application of equations (2) and (3) for modeling earthquake distributions.

**Discussion**

The coefficient *b*,

$$b = \frac{s}{\alpha + 1} \log_{10} e, \qquad (12)$$

analogous to the Gutenberg and Richter *b-value*, was calculated for each area, considered in this work, using the coefficients *s* and *α* of the function (2) approximating the observed cumulative distribution of earthquakes, Figure 2.

The table below summarizes this calculation. This Table also shows the expected magnitudes of earthquakes taken from the probability distribution functions of these areas, Figure 2.

**Table**

| Area<br>Quantity | Central Atlantic | Canary Islands | Sea of Japan | Magellan Mountains |
|---|---|---|---|---|
| Coefficient *b*, calculated by formula (12) | *1.075* | *1.5* | *1.76* | *1.8* |
| The expected magnitude *m* | *3.7* | *2.5* | *4.8* | *4.4* |

The Gutenberg-Richter *b-value* calculated for the Magellan Mountains region is *1.3*. The coefficient *b*, calculated for this region by formula (12), is *1.8*. This difference can be explained by the fact that the exponential function approximates the observed distribution for magnitudes *m > 4* only, while the coefficient *b* was calculated by the formula (12) with coefficients *s* and *α* obtained for the whole curve, approximating a range of earthquake magnitudes, from *1* to *8*. The excellent fit between the observed data and the solution of the generalized logistic equation, presented in this work, indicates that this equation can become a new starting point in understanding the physics of earthquakes and constructing statistical models explaining the spacial and temporal distribution of earthquakes.

**Data**

All the data used in this work were taken from the U.S. Geological Survey Earthquake Data Base website: www.earthquake.USGS.gov/earthquakes/eqarchives/epic/epic_rest.php

**Conclusion**

The Gutenberg-Richter frequency-magnitude formula was derived as an asymptotic case for large magnitudes from the solution of the generalized logistic equation.

Approximation of the observed cumulative distribution of earthquakes was given for four different geological provinces: in the Central Atlantic (40N-25N, 5W-35W), Canary Islands, Magellan Mountains (20N-9S, 148E-170E), and the Sea of Japan. The solution of the generalized logistic equation showed an excellent fit between the theoretical and observed data for magnitudes *1<m<9*.

The new approach to the modeling and study of earthquakes, presented in this work, can provide us with new insight into the physics of earthquakes.

**Acknowlegements**

Authors are grateful to Prof. V.I. Chebotarev for his comments and suggestions which helped to improve the quality of this work.